\documentclass[11pt,a4paper]{article}
\usepackage[margin=2.5cm]{geometry}

\usepackage[utf8]{inputenc}
\usepackage[T1]{fontenc}

\usepackage{amsmath,amssymb,amsfonts}

\usepackage{graphicx}
\usepackage{subfig}
\usepackage{array}
\usepackage{booktabs}
\usepackage{float}

\usepackage{xcolor}
\usepackage{url}
\usepackage{hyperref}

\title{\textbf{Deep-learning surrogate model for rapid
quantum-noise computation in dual-pump fiber-optic
phase-sensitive amplifiers}}

\author{
Yousra Bouasria$^{1,*}$,
Debanuj Chatterjee$^{2,*}$,
Abdessamad Belfakir$^{3}$,
Debabrota Basu$^{4}$\\[1em]
\small $^{1}$Independent Researcher, Paris, France\\
\small $^{2}$Independent Researcher, Kolkata, India\\
\small $^{3}$UM6P Vanguard Center,
Mohammed VI Polytechnic University (UM6P), Morocco\\
\small $^{4}$University of Lille, Inria, CNRS,
Centrale Lille, UMR 9189--CRIStAL, Lille, France\\[0.5em]
\small Corresponding author:
\texttt{bouasriayousra@gmail.com}\\
\small $^{*}$These authors contributed equally to this work.
}

\date{}

\begin{document}

\maketitle

\begin{abstract}
Fiber-optic parametric amplifiers in a phase-sensitive amplifier\,(PSA) configuration are well known for their low noise properties. Quantum mechanics dictates that these amplifiers can have a 0 dB noise figure\,(NF) (ratio of signal-to-noise ratio at the input and output of the amplifier) under certain conditions. However, for a dual-pump configuration, when the system is highly nonlinear, fully quantum calculations of NF become non-trivial. In such cases, semi-quantum models of NF evaluation exists, however, they are often complicated and resource intensive. To simplify that, in this work, first we use a semi-quantum NF calculation framework for the generation of numerically simulated input-output dataset of NF over a large parameter spaces. Then we use the data to train an optimized deep neural network\,(DNN) for rapid NF prediction.
Utilizing the DNN, we demonstrate about 400 times reduction in the computation time compared to the standard semi-quantum NF evaluation approach with a mean absolute error\,(MAE) of 0.0015\,dB.
Using clustering we demonstrate a rapid exploration of the parameter space with respect to high gain (20 dB) and low NF (0.3 dB) targets and retrieve associated design parameters. These results illustrate the efficacy of machine learning-based approaches for rapid search of high-dimensional parameter spaces especially in compute intensive photonic systems.    
\end{abstract}

\par

\section{Introduction}
Phase sensitive amplification (PSA) has been extensively investigated in recent years because of its potential to provide noiseless optical amplification and generate non-classical squeezed states of light\,\cite{Tong2012UltralowNoisePSA,Andrekson2020FiberPSA}. Based on nonlinear optical interactions, PSAs can be implemented in media having second order nonlinear properties ($\chi^{(2)}$), such as periodically poled lithium niobate waveguides \cite{Umeki2013InlinePSA}, as well as those with third order nonlinear properties ($\chi^{(3)}$) such as highly nonlinear fibers\,(HNLF)\,\cite{McKinstrie2004PSAFiber, Slavik2010Regenerator, labidi2018phase} and dispersion-engineered nonlinear waveguides \cite{Neo2013PSAChalcogenide, zhang2014phase}. 
Due to their ability to selectively amplify input signals with certain input phases, PSAs have been considered for various applications in optical communications\,\cite{Tong2012UltralowNoisePSA,Andrekson2020FiberPSA}, microwave photonics\,\cite{chatterjee2020investigation, zhao2020enhanced}, quantum information science\,\cite{Takanashi2020AllOpticalPhaseSensitiveDetection,Liu2021PhaseManipulatedEntangledState,shringarpure2019generating,erkilicc2026unified}, 
and photonic signal processing\,\cite{Slavik2010Regenerator,DaRos2013QPSKtoBPSK,chatterjee2022optimization}. 
A key feature of phase sensitive amplification is its ability to achieve, ideally, a quantum-limited noise figure (NF) of 0 dB, corresponding to amplification without degradation of the input signal-to-noise ratio (SNR), in a shot-noise limited signal. In contrast, phase insensitive amplification (PIA) necessarily introduces excess quantum noise, resulting in a minimum NF of about 3 dB in the high-gain regime. This fundamental difference arises because PIA must preserve the quantum commutation relations, which requires the introduction of vacuum fluctuations that are amplified together with the signal and ultimately degrade the detected SNR \cite{Caves1982QuantumLimits}.

Phase sensitive amplifiers based on HNLFs can be operated in single and dual-pump configurations. A dual-pump configuration with degenerate signal-idler is often preferred due to its larger maximum PSA gain\,\cite{Xie2015MultiWavePSA}. Although dual-pump fiber-optic PSAs can theoretically operate at the 0 dB NF limit, in practice when the nonlinearity of the fiber becomes large, multiple parasitic four-wave mixing\,(FWM) processes may occur in addition to the desired phase sensitive interaction. These processes generate additional sidebands that couple with the signal and modify its properties\,\cite{Xie2015MultiWavePSA, Chatterjee2021SevenWaveModel, Marhic2008FiberOPA}. Moreover, the generated sidebands introduce additional input ports for vacuum fluctuations, which can degrade the NF. Previous theoretical and numerical studies based on different approaches have shown that such parasitic nonlinear interactions can lead to deviations from the ideal 0 dB NF limit, depending on the fiber parameters and wavelength allocation \cite{Bouasria2020NoiseFigureMultiWave, mckinstrie2004quantum, chatterjee2022optimization, Bouasria2020GeneralisedNF, McKinstrie2005QuantumNoiseParametricProcesses}, to name a few.

In the literature, accurate evaluation of noise in PSAs require a quantum-mechanical treatment of the interacting optical modes. However, fully quantum calculations become increasingly challenging in highly nonlinear dual-pump configurations because of the growing number of coupled modes, nonlinear interactions, and constraints imposed by quantum commutation relations. To address this difficulty, a practical semi-quantum numerical approach has been developed \cite{Inoue2019MultipleFWMNoise}, combining classical nonlinear propagation of the pumps and high-order pumps\,(HOP) with quantum calculations for the noise on the signal, under realistic operating conditions including the power depletion of the pumps. 

Nevertheless, such semi-quantum frameworks remain computationally demanding, because it requires solving the pump and HOP propagation in the fiber, numerically.
Such resource intensive methods act as bottlenecks for amplifier design, requiring search over a large PSA parameter spaces.

Machine learning (ML) provides a promising alternative for accelerating computationally intensive physical modeling tasks by learning complex relationships directly from simulation data \cite{hager2020physics, Willard2022ScientificKnowledgeML, Karniadakis2021PhysicsInformedML, Wang2025MLFiberChannels, Boscolo2023NonlinearReshaping, Raissi2019PINN, Jiang2022PINNFiberDynamics, Song2022PhysicsInformedNeuralOperator}. ML-based surrogate models %
are not directly constrained by discretization-step size, and hence once trained, a surrogate model can provide rapid exploration of the parameter space covered by the training data, while retaining the underlying physics. These capabilities are essential for developing efficient simulation strategies and supporting the design of next-generation photonic systems\,\cite{Hirschman2026DeepLearningChi2, Genty2021MLUltrafastPhotonics}.

While previous works investigated ML-based models for learning wave propagation dynamics through nonlinear fibers\,\cite{ermolaev2022data,boscolo2020artificial, boscolo2021modelling, Boscolo2023NonlinearReshaping}, surrogate development for quantum noise in PSAs remain relatively unexplored. Thus, to enable efficient exploration of the parameter space for fiber PSA design, we develop an optimized deep neural network\,(DNN) model trained on numerically generated input-output datasets, obtained from a semi-quantum NF calculation framework to accurately predict the maximum gain and NF of a dual-pump PSA. The DNN model reduces computation time significantly compared to standard approaches, while maintaining a high NF prediction accuracy.

This paper is organized as follows. In Section\,\ref{sec:Phys} we introduce a theoretical semi-quantum framework based on multiwave model-based generation of the pump and HOP data as provided in\,\cite{Inoue2019MultipleFWMNoise}. Then in Section\,\ref{sec:DL} we introduce the deep learning\,(DL) architecture for the developed surrogate model. After that in Section\,\ref{sec:res} we outline the obtained results with our surrogate model. In Section\,\ref{sec:cluster} we make an exploration of the parameter space, and retrieve useful parameters that lead to a high gain as well as a low noise figure, as required in a high-performance amplifier design. Finally in Section\,\ref{sec:con} we draw the conclusion and outline some perspectives of the work.

\section{Semi-Quantum Noise Figure Model}\label{sec:Phys}
\begin{figure}[htb!]
   \centering
   \includegraphics[width=\textwidth]{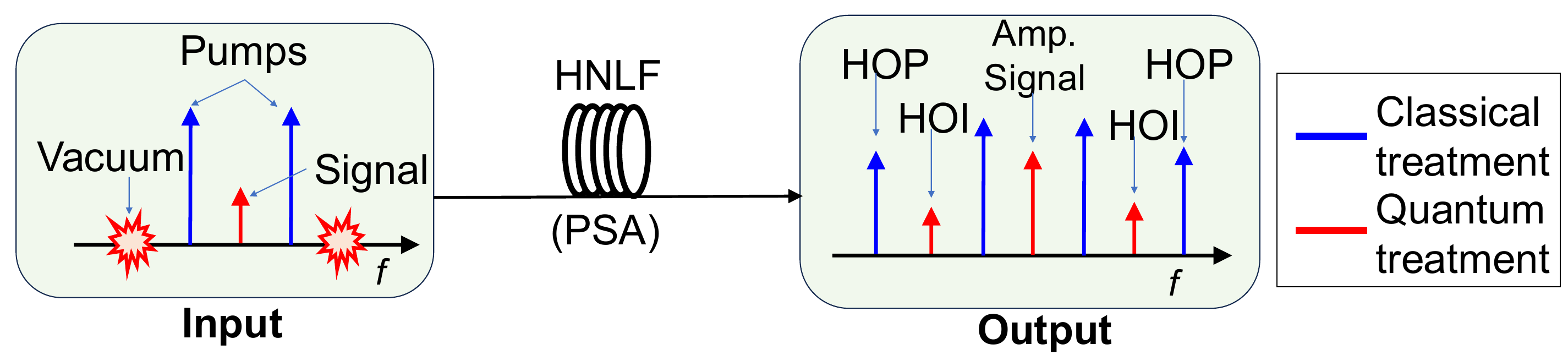}
    \caption{Illustration of input-output frequency configuration of a dual-pump phase sensitive amplifier\,(PSA), with degenerate signal/idler. In our model, waves treated classically are shown in blue and waves treated quantum mechanically are shown in red. HOP : high-order pump, HOI : high-order idler, HNLF : highly nonlinear fiber, Amp. : Amplified. (not to scale)}\label{fig_config}
\end{figure}

\begin{figure}[H]
   \centering
   \includegraphics[width=\textwidth]{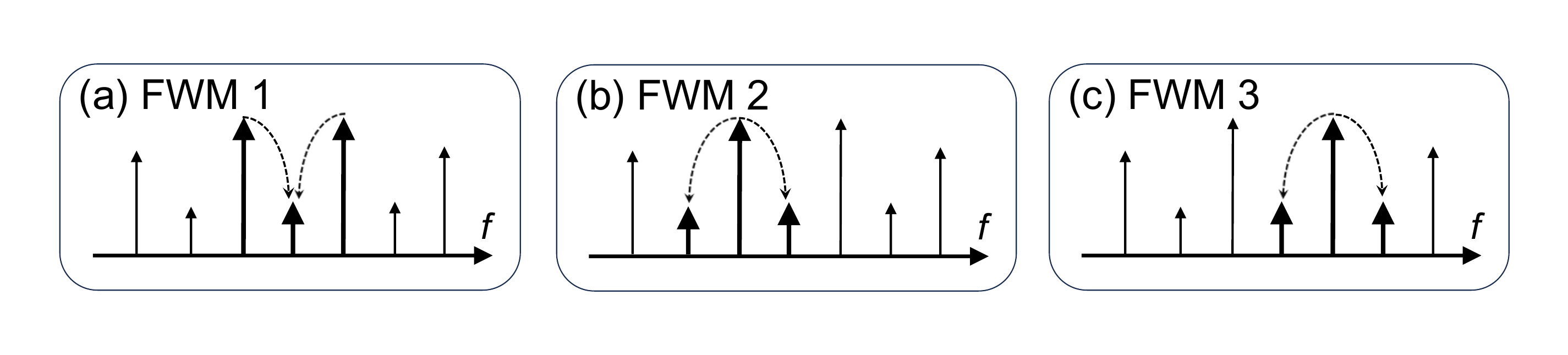}
    \caption{Illustration of different four wave mixing\,(FWM) processes consisting of at least two pump photons and involving the signal. We have not shown frequency conversion FWM processes. The dotted arrows show the direction of photon conversion, and the involved waves are in bold for visual clarity. (not to scale)}\label{fig_fwm}
\end{figure}

In this section we introduce a semi-quantum NF calculation model for a dual-pump fiber PSA with degenerate signal/idler (which we will refer to as signal throughout this manuscript). The system configuration with the different waves are shown in Fig.\,\ref{fig_config}. The model use a framework, where the evolution of the signal and high-order idlers\,(HOI) are treated quantum mechanically, whereas the pumps and high order pumps\,(HOP) are treated classically\,\cite{Inoue2019MultipleFWMNoise}. The classical evolution of the pumps and HOPs along the fiber are solved numerically to find their complex amplitude at each small segment of length $\text{d}z$ within the fiber. This model is adapted from \cite{Inoue2019MultipleFWMNoise}, where a multiwave model is used for solving the pump and HOP waves. 
A dual-pump PSA with degenerate signal consists of a signal wave at frequency $f_\text{s}$ (wavelength $\lambda_\text{s}$) and two strong pumps at frequencies $f_{\text{P1}}$ and $f_{\text{P2}}$ (wavelengths $\lambda_{\text{P1}}$ and $\lambda_{\text{P2}}$) at the input\,(see Fig.\,\ref{fig_config}). The interacting waves satisfy the energy-conservation relation:
\begin{equation}
2f_\text{s}=f_{\text{P1}}+f_{\text{P2}}.
\end{equation}
In a quantum-mechanical picture, one photon from each pump is converted into two signal photons through a FWM interaction\,[see process FWM 1 in Fig.\,\ref{fig_fwm}\,(a)], resulting in phase-sensitive amplification. If we consider this single FWM process, the SNR undergoes no degradation, and the quantum NF therefore should be $0$~dB. In a HNLF however, the same input waves also generate additional spectral components such as high-order idlers\,(HOI) through other FWM processes\,[see processes FWM 2 and FWM 3 in Fig.\,\ref{fig_fwm}\,(b,c)] which individually correspond to phase-insensitive amplification processes (since there are no HOIs at the input). These parasitic FWM processes couple the input vacuum fluctuations to the signal mode and hence degrade the signal NF from its ideal value of $0$~dB. In this model we only consider processes FWM 1, 2 and 3, since those are the ones involving signal photons and at least two pump photons. We have not considered frequency conversion FWM processes since they do not contribute in NF degradation\cite{mckinstrie2004quantum, Bouasria2020GeneralisedNF}.Also FWM processes involving HOPs were assumed to be sufficiently decoupled from the signal\,\cite{Chatterjee2021SevenWaveModel}, and hence not considered in the quantum-mechanical calculations\, \cite{Bouasria2020NoiseFigureMultiWave}.
In this framework, the signal and idler waves are represented by quantum-mechanical annihilation and creation operators. The pump evolution is obtained by numerically solving for the two pumps and HOPs along the fiber. Quantum operators are then used to describe the evolution of the signal and the two idler modes and to evaluate the associated NF. To account for variations of the pump powers and phase mismatch along the propagation direction, the fiber is divided into a large number of short segments. Within each segment, the pump fields are assumed to be approximately constant. The signal and idler operators are propagated through each segment using the Heisenberg equations of motion derived from the interaction Hamiltonian. The input-output relation of one segment is written in matrix form, and the transfer matrix of the complete fiber is obtained by multiplying the matrices of all segments. The signal operator at the fiber output can be expressed as a linear combination of the input signal, its conjugate, and the two idler vacuum operators as\,\cite{Inoue2019MultipleFWMNoise}:
\begin{equation}\label{eq1}
\begin{aligned}
\hat{a}_{s,\mathrm{out}}
=
C_1\hat{a}_{s,\mathrm{in}}
+
C_2\hat{a}_{s,\mathrm{in}}^{\dagger}
+
C_3\hat{a}_{i1,\mathrm{in}}
+
C_4\hat{a}_{i1,\mathrm{in}}^{\dagger}
+ \\
C_5\hat{a}_{i2,\mathrm{in}}
+
C_6\hat{a}_{i2,\mathrm{in}}^{\dagger},
\end{aligned}
\end{equation}
where the coefficients $C_1,\ldots,C_6$ are obtained from the total transfer matrix. After scanning the input signal phase, the maximum signal gain is written as:
\begin{equation}\label{eq:gmax}
G_{\mathrm{max}}=\left(|C_1|+|C_2|\right)^2.
\end{equation}
The terms involving $C_3$--$C_6$ represent the coupling of the idler vacuum modes to the signal output and therefore quantify the excess noise generated by the HOIs and pumps. By evaluating the quantum photon-number variance, the NF expression, in linear scale, is given as\,\cite{Inoue2019MultipleFWMNoise}:
\begin{equation}\label{eq:NF}
\mathrm{NF}
=
1+
\frac{
|C_3|^2+|C_4|^2+|C_5|^2+|C_6|^2
}{G_\mathrm{max}}.
\end{equation}
Note, that if only FWM 1 is considered, the coefficients $C_3$--$C_6$ vanish in Eq.\,\ref{eq1} and the ideal $0$~dB NF is recovered. Thus the strength of the degradation of the NF, is governed by the  parasitic FWM processes (FWM 2 and 3), which in turn depends on the system parameters, such as pump powers, signal-pump1 wavelength separation $\lambda_{sp}$, fiber nonlinear coefficient $\gamma$,  dispersion slope of the fiber $D_{cc}$, and zero-dispersion wavelength of the fiber $\lambda_{ZDW}$. In our NF computation framework, we use these parameters at the input of the DNN.

\subsection{Pumps and HOPs Evolution}
\begin{figure}[htb!]
   \centering
    \includegraphics[width=\textwidth]{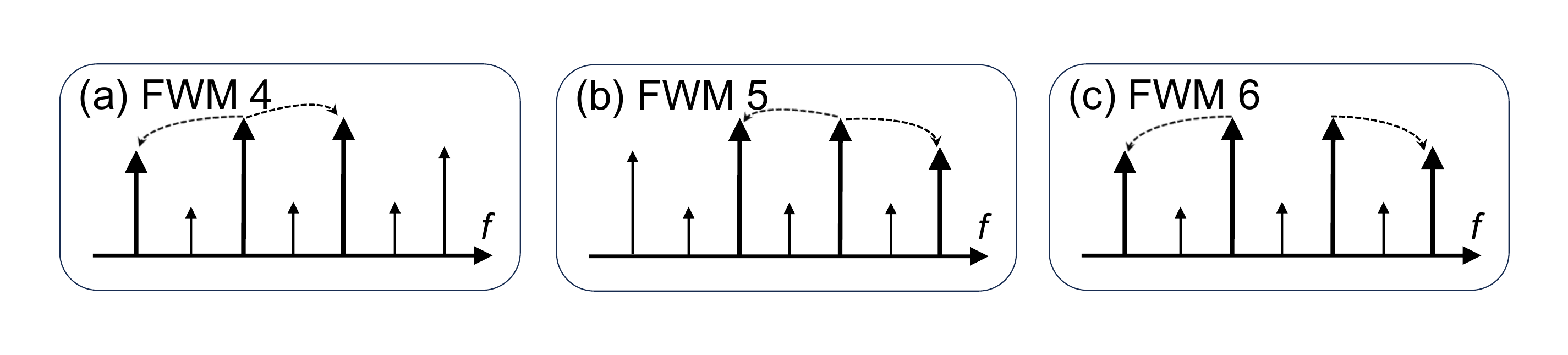}
    \caption{Illustration of different FWM processes involving the pumps and the HOPs. The dotted arrows show the direction of photon conversion, and the involved waves are in bold for visual clarity. (not to scale)}\label{fig_fwmP}
\end{figure}

A multiwave model is used to numerically evaluate the propagation of the two pumps and the HOPs through the fiber. The propagation equation of the slowly varying complex amplitudes of the pumps ($A_{\text{P1}}$ and $A_{\text{P2}}$) and the two HOPs ($A_{\text{HOP1}}$ and $A_{\text{HOP2}}$) along $z$ are given as\,\cite{Inoue2019MultipleFWMNoise}: 
\begin{align}
\frac{dA_{\mathrm{P1}}}{dz}
={}& 
2i\gamma A_{\mathrm{HOP1}}A_{\mathrm{P2}}A_{\mathrm{P1}}^{*}
e^{-i\Delta\beta_{\mathrm{1}}z}
+
i\gamma A_{\mathrm{P2}}^{2}A_{\mathrm{HOP2}}^{*}
e^{i\Delta\beta_{\mathrm{2}}z},
\label{eq:pump_depletion_a}
\\[4pt]
\frac{dA_{\mathrm{P2}}}{dz}
=&
2i\gamma A_{\mathrm{HOP2}}A_{\mathrm{P1}}A_{\mathrm{P2}}^{*}
e^{-i\Delta\beta_{\mathrm{2}}z}
+
i\gamma A_{\mathrm{P1}}^{2}A_{\mathrm{HOP1}}^{*}
e^{i\Delta\beta_{\mathrm{1}}z},
\label{eq:pump_depletion_b}
\\[4pt]
\frac{dA_{\mathrm{HOP1}}}{dz}
&=
i\gamma A_{\mathrm{P1}}^{2}A_{\mathrm{P2}}^{*}
e^{i\Delta\beta_{\mathrm{1}}z},
\label{eq:pump_depletion_c}
\\[4pt]
\frac{dA_{\mathrm{HOP2}}}{dz}
&=
i\gamma A_{\mathrm{P2}}^{2}A_{\mathrm{P1}}^{*}
e^{i\Delta\beta_{\mathrm{2}}z},
\label{eq:pump_depletion_d}
\end{align}
where the total effective phase mismatch (linear and nonlinear) $\Delta\beta_{\mathrm{1}}$ and $\Delta\beta_{\mathrm{2}}$ are given as:
\begin{equation}
\begin{split}
&\Delta\beta_{\mathrm{1,2}}
=2\beta\!\left(f_{\mathrm{P1,2}}\right)
-\beta\!\left(f_{\mathrm{P2,1}}\right)
-\beta\!\left(f_{\mathrm{HOP1,2}}\right)
- \gamma\left(
2\left|A_{\mathrm{P1,2}}\right|^{2}
-\left|A_{\mathrm{P2,1}}\right|^{2}
\right),
\end{split}
\end{equation}
where $f_{\mathrm{P1}}$ and $f_{\mathrm{P2}}$ are the frequencies of the pumps and $f_{\mathrm{HOP1}}$ and $f_{\mathrm{HOP2}}$ are those of the HOPs. $\beta(f)$ is the linear propagation constant at frequency $f$. These evolution equations [Eqs.\,(\ref{eq:pump_depletion_a}--\ref{eq:pump_depletion_d})] account for FWM processes with at least three pump photons [FWM 4 and 5 as shown in Figs.\,\ref{fig_fwmP}\,(a,b)]. FWM 6, shown in Fig.\,\ref{fig_fwmP}\,(c) however although a legitimate FWM process, is not included in this model, since it involves only two pump photons, and hence is assumed to be weaker than processes with three pump photons. 
\section{Deep-Learning Architecture}\label{sec:DL}
\begin{figure}[htb!]
    \centering
    \includegraphics[width=0.9\textwidth]{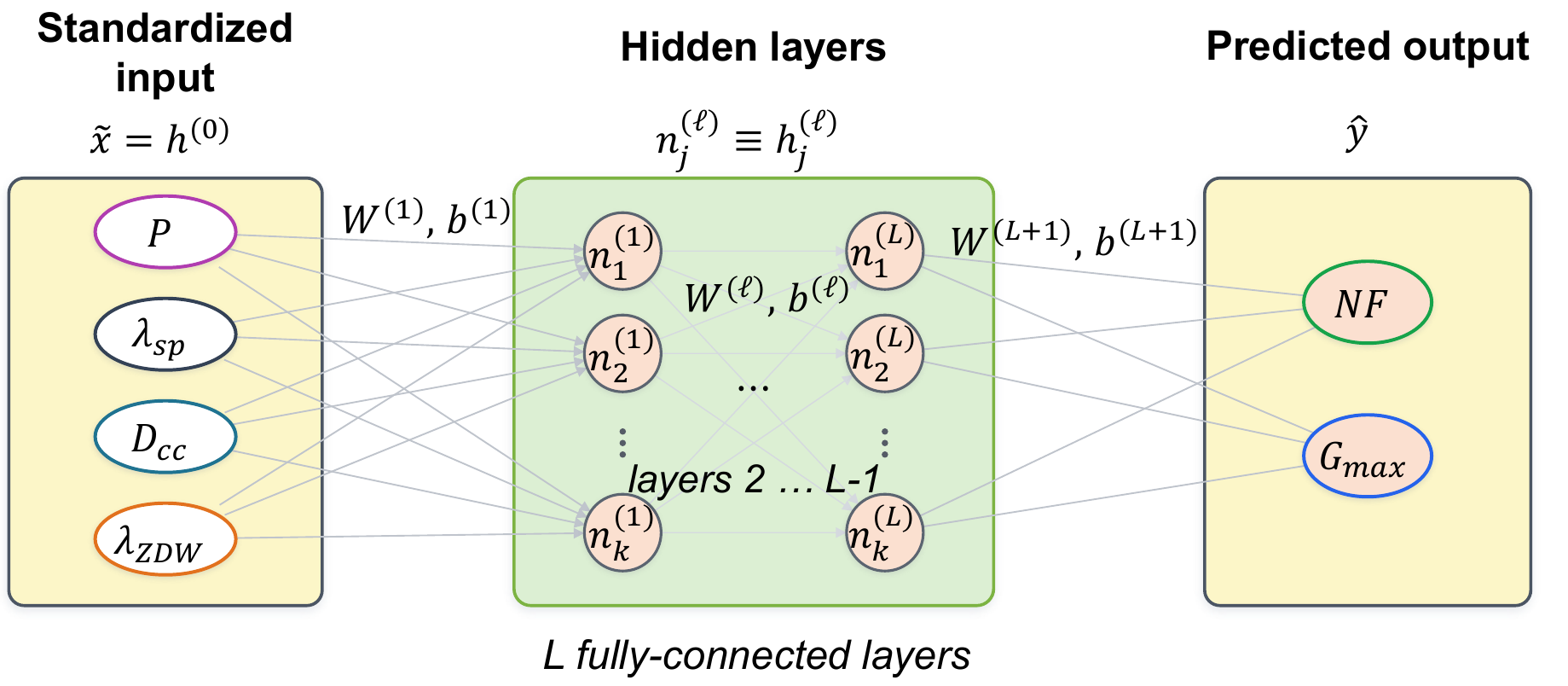}
    \caption{Architecture of the fully connected deep-learning surrogate model. The network maps the four standardized physical features, collected in the standardized input vector $\tilde{x}=h^{(0)}$, onto the predicted output $\hat{y}$ through $L$ fully connected hidden layers. 
    A network depth of $L=6$ is considered.
    $W^{(\ell)}$ is the weight matrix, $b^{(\ell)}$ the bias vector, and $h^{(\ell)}$ the activation vector of layer $\ell$. Each node $n_{j}^{(\ell)}$ is the $j$-th neuron (hidden unit) of hidden layer $\ell$; its value equals the $j$-th component of the activation vector, i.e.\ $n_{j}^{(\ell)}\equiv h_{j}^{(\ell)}$ for $j=1,\dots,k$, where $k$ is the number of hidden units in the layer.}
\label{fig1}
\end{figure}

The semi-quantum noise figure model introduced in the previous section is quite resource intensive, it poses challenges in the efficient design of a dual-pump PSA. To circumvent that, in this section, we propose the deep learning-based surrogate architecture to approximate the nonlinear relationship between the physical input parameters of the dual-pump fiber PSA and its noise figure and maximum gain. The surrogate model is implemented as a DNN composed of fully connected layers, as illustrated in Fig.~\ref{fig1}.  
Each hidden layer transforms the activation vector of the preceding layer according to : 
\begin{equation}
h^{(\ell)}
=
\varphi\!\left(
W^{(\ell)}h^{(\ell-1)}
+
b^{(\ell)}
\right),
\end{equation}
where \(W^{(\ell)}\) and \(b^{(\ell)}\) denote the weight matrix and bias vector of layer \(\ell\), respectively, and \(\varphi\) is the nonlinear activation function. The output layer applies a linear activation to directly estimate the two continuous targets.
The trainable weights and biases, collectively denoted by
$\theta$, are determined by minimizing a loss function
$\mathcal{L}$ over the training data using gradient-based optimization,
with the required gradients computed through backpropagation. Once trained, the surrogate model simultaneously predicts
the noise figure and the maximum phase sensitive gain through a single
forward pass.
In the following subsections,
we describe the numerical generation of the training datasets using the described model.
In the following we provide a detailed discussion on the DNN based framework.
\subsection{Data Generation}
The dataset for the pumps and the HOPs are generated numerically using the multiwave model. For each operating point (set of 4 input parameters), the model receives the pump-signal wavelength separation $\lambda_{sp}$, dispersion slope $D_{cc}$, zero-dispersion wavelength $\lambda_{\mathrm{ZDW}}$ of the fiber, and input power of each pump $P$. The fiber length and nonlinear coefficient are fixed to $\gamma=12$ (W.km)$^{-1}$ and $Z=300$ m. Since the nonlinear dynamics of the system is primarily dependent on the nonlinear phase $\gamma PZ$ as a whole, only one of the three terms can be varied for most practical purposes. In our case we vary the pump power $P$. $P$ was varied between 19 dBm and 25 dBm, making the nonlinear phase vary between 0.3 rad and 1.1 rad. The signal wavelength is fixed at 1550 nm.
From each simulation run, $C_1...C_6$ was obtained from the pump and HOP evolution data, and the maximum gain and NF of the signal was computed using Eqs.\,(\ref{eq:gmax}) and (\ref{eq:NF}) respectively\,\cite{Inoue2019MultipleFWMNoise}. A Runge-Kutta 4 integration method was used to solve propagation equations for pumps and HOPs. 300000 instances were simulated, where the parameter values were drawn from an uniform random distribution within a maximum and minimum range as shown in Table\,\ref{tab}. A separate independent dataset of 50000 instances was also generated for external testing of the model. 
\begin{table}[H]
    \centering 
    \caption{Maximum and minimum range of input parameters for the generated dataset.}
    \label{tab}
    \begin{tabular}{|m{19em}|m{8em}|m{8em}|}
    \hline\centering
        \textbf{Parameter} & \textbf{Minimum}& \textbf{Maximum}  \\ \hline \hline
        Pump power ($P$)&19 dBm&25 dBm\\ \hline
        Pump-signal wavelength separation ($\lambda_{sp}$)&2 nm&25 nm\\ \hline
        Fiber zero-dispersion wavelength ($\lambda_{\text{ZDW}}$)&1500 nm&1600 nm\\ \hline
        Dispersion slope ($D_{cc}$)&0.01 ps/(nm$^2$km)&0.07 ps/(nm$^2$km)\\ \hline
    \end{tabular}
\end{table}
\subsection{Machine learning framework}\label{sec:ml_framework}
A feed-forward neural network is trained to approximate the responses of the NF model. The input and output vectors are defined as $\Tilde{x}=[\lambda_{sp},D_{cc}, \lambda_{\mathrm{ZDW}}, P]^{\mathrm{T}}$, and $\hat{y}=[\mathrm{NF},G_{\max}]^{\mathrm{T}}$.
The forward mapping is approximated by $\hat{y}=f_{\theta}(x)$, $\theta$ contains the trainable weights and biases of the network, and $f_{\theta}$ denotes the complete surrogate mapping.
The dataset of each model is divided into 80$\%$ training and 20$\%$ validation sets.
In order to efficiently implement a DNN framework, the input features and targets of the model are standardized to comparable scales. The same preprocessing is applied to the validation and test datasets.
The developed surrogate model is DNN, composed of fully connected layers.
The DNN have six hidden layers. The layer widths are listed in Table~\ref{tab:ml_hyperparameters}, which were obtained by hyper-parameter optimization runs. 
Each hidden unit in the hidden layers applies an affine transformation followed by a LeakyReLU activation function with a negative slope of 0.01. 
The DNN surrogate model minimizes the mean absolute error (also called $L1$ loss) evaluated with the standardized dataset. The absolute residuals are first averaged over the two outputs (NF and $G_{\mathrm{max}}$) and the resulting sample losses are then averaged over the mini-batch.
The network parameters are estimated through backpropagation using PyTorch automatic differentiation \cite{paszke2019pytorch}. Parameter updates are performed using the AdamW optimizer, 
with a cosine learning
rate annealing.
Each network is trained for a maximum of 500 epochs, with early stopping after 60 epochs when the validation loss does not improve. The network checkpoint yielding the minimum validation loss is retained for subsequent evaluation. 
Final generalization performance is assessed using an independent external test dataset for each model. 
The hyperparameters used in the preceding training were selected  through a separate optimization procedure implemented using Ray Tune optimization\,\cite{liaw2018tune} (see Appendix).
After completing the hyperparameter optimization, the configuration achieving the lowest validation loss across all trails and epochs is saved, which make the final optimized DNN.
All relevant hyperparameters are provided in Appendix.
\section{NF and Gain Prediction Results}\label{sec:res}
Building on the machine-learning framework presented in
Subsection~\ref{sec:ml_framework}, this section assesses the
predictive performance of the trained surrogate model.
\begin{figure}[htb!]
    \centering
    \includegraphics[width=0.85\textwidth]{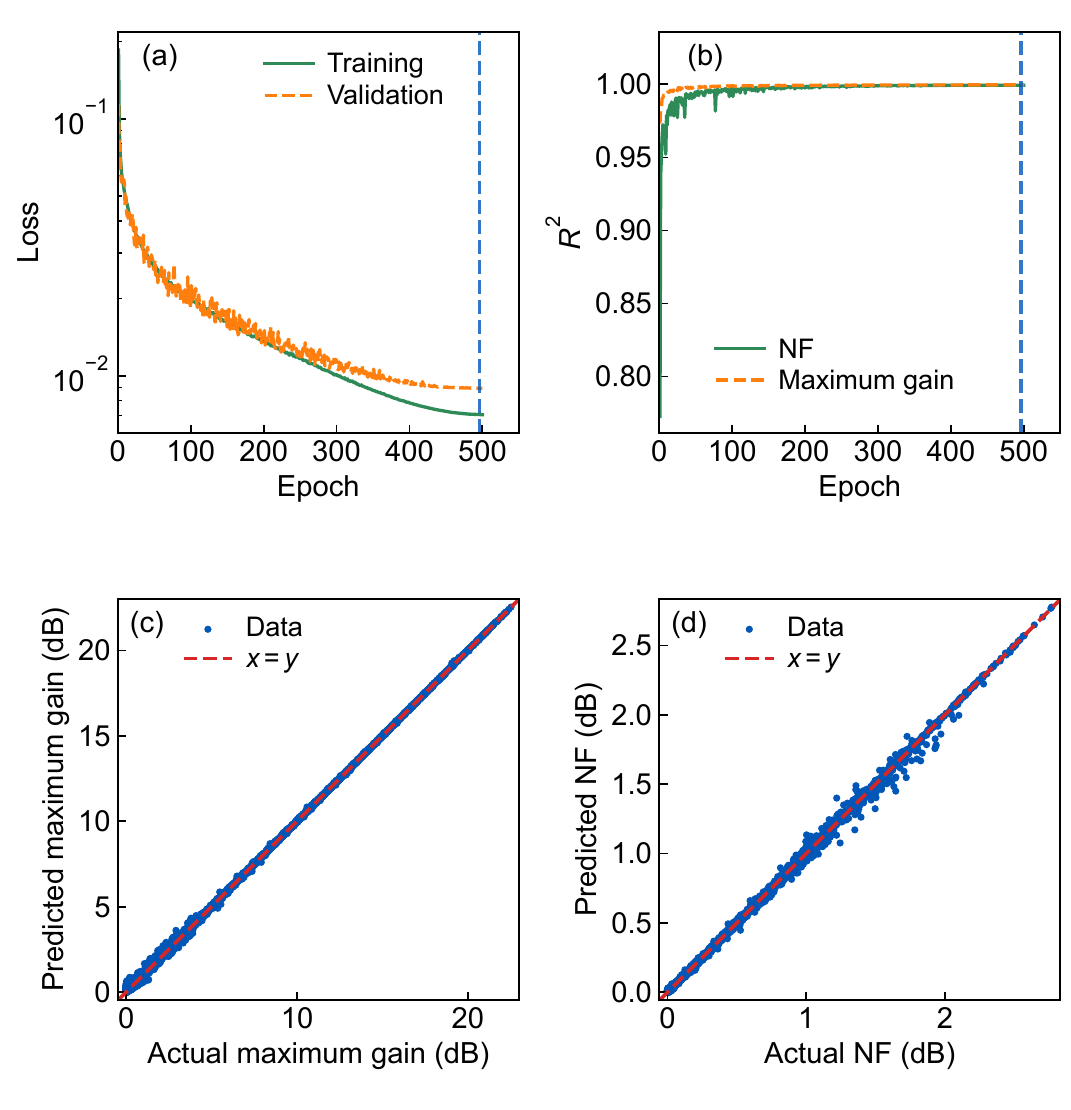}
    \caption{Performance of the trained surrogate model.
    (a) Training and validation loss as a function of the training epoch.
    (b) Coefficient of determination $R^2$ for the noise figure $\mathrm{NF}$
    and maximum gain $G_{\max}$ during training.
    (c) Predicted versus actual maximum gain on an unseen external dataset.
    (d) Predicted versus actual NF on an unseen external dataset.
    The dashed red line in (c) and (d) represents the ideal prediction,
    $y=x$.}
    \label{fig:NNModel_I}
\end{figure}
Fig.~\ref{fig:NNModel_I}(a) shows the evolution of the
training and validation losses over 500 epochs. The green (training) and orange (validation) curves show the evolution of the loss on a logarithmic scale as a function of the epoch, for the two outputs of the DNN. The losses decrease rapidly during the initial epochs and reach approximately 0.007 and 0.009  for training and validation, respectively, at the end of training.
Both training and validation curves show stable convergence, with the validation loss approaching a plateau and no clear evidence of overfitting.
Correspondingly, in Fig.~\ref{fig:NNModel_I}(b) we present the coefficient of
determination $R^2$ for NF and maximum gain. Both coefficients increase
rapidly during the early epochs,
reaching approximately 0.9997 for maximum gain and 0.9994 for NF by the end of training. This
indicates that the trained network captures most of the
variation of both the target quantities.

To assess the generalization capability of the DNN model,
Figs.~\ref{fig:NNModel_I}(c) and~(d) compare the predicted and
actual values of maximum gain and NF, respectively, for an
external dataset of 50000 samples. The predictions closely follow the $x=y$ line over the investigated ranges, extending to approximately 22~dB for $G_{\max}$ and 2.8~dB for NF. A slightly larger deviation from the ideal $y=x$ relation is observed particularly for NF values between approximately 1 and 2~dB.

On this external dataset, we found that most absolute prediction errors for
the maximum gain remain below approximately 0.2~dB, while for the NF, it is below 0.1~dB. The mean absolute error (MAE) achieves 0.0398~dB for maximum gain and 0.0015~dB for NF. These results are consistent
with the close agreement between the predicted and actual
values shown in Figs.~\ref{fig:NNModel_I}(c) and~(d), and show that the model maintains small absolute prediction errors over most of the external dataset. Overall, the results support the
ability of DNN to generalize to unseen datasets within the investigated parameter domain.

In order to highlight the advantage of our DNN-based computational framework, we compared the mean calculation time of NF using the optimized DNN framework, for a set of input parameters. We then compared these computation times with that of the standard semi-quantum approach. To make the comparison fair, we made the calculations in the same hardware, over 500 different runs. The compute time reduced from 80 ms to about 0.2 ms as we moved from the standard to the DNN approach, resulting in a speedup of about 400 times.

\section{NF-Gain Clustering}\label{sec:cluster}
\begin{figure}[htb!]
    \centering
\includegraphics[width=0.7\textwidth]{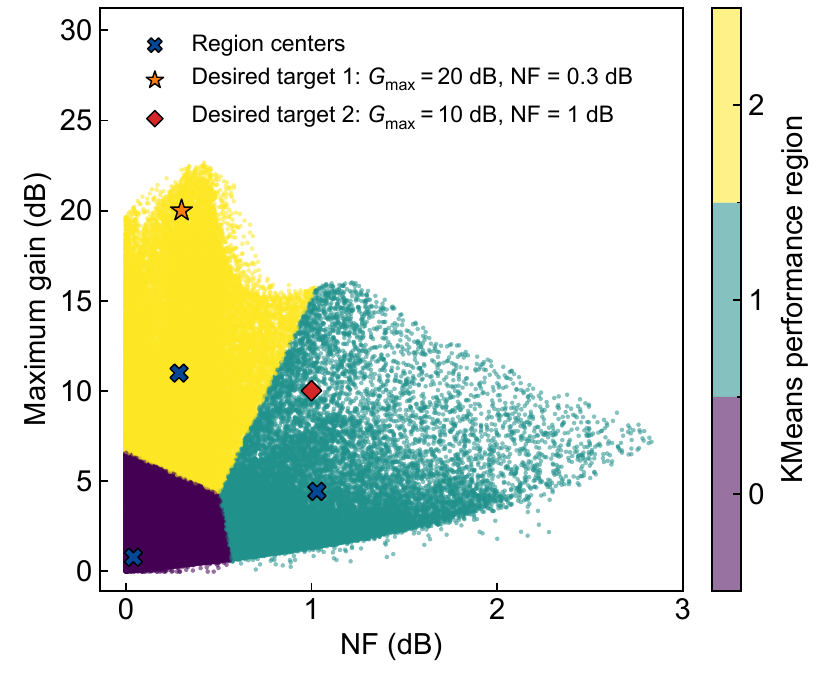}
    \caption{K-means partitioning of the NF--gain space, where the cross markers indicate the cluster centers and the star and diamond markers represent desired target points 1 and 2, respectively.}
\label{fig:kmeans_performance_space}
\end{figure}

Beyond the prediction of NF and maximum gain, presented in Section\,\ref{sec:res}, in this section we aim to identify the input conditions associated with specific performance targets. As shown in Fig.~\ref{fig:kmeans_performance_space} we plot the NF-gain distributions of the dataset. 
In order to group samples with similar output
characteristics in the NF-gain space, we perform $K$-means clustering of the data points. We investigate the input parameters of the group of data-points near a specified target region.
Before clustering, $\mathrm{NF}$ and $G_{\max}$ are standardized so that each contributes to the distance metric in proportion to its standard deviation. The resulting regions are displayed in the original output units. Region~0 (in purple) contains low-NF, low-gain samples, with a center near $(\mathrm{NF},G_{\max})\approx(0.04,\,0.75)$~dB; region~1 (in teal) covers intermediate operating conditions, centered near $(1.02,\,4.41)$~dB; region~2 (in yellow) contains the high-gain and low noise samples, centered near $(0.28,\,10.98)$~dB. 
\begin{figure}[htb!]
    \centering
    \includegraphics[width=0.8\textwidth]{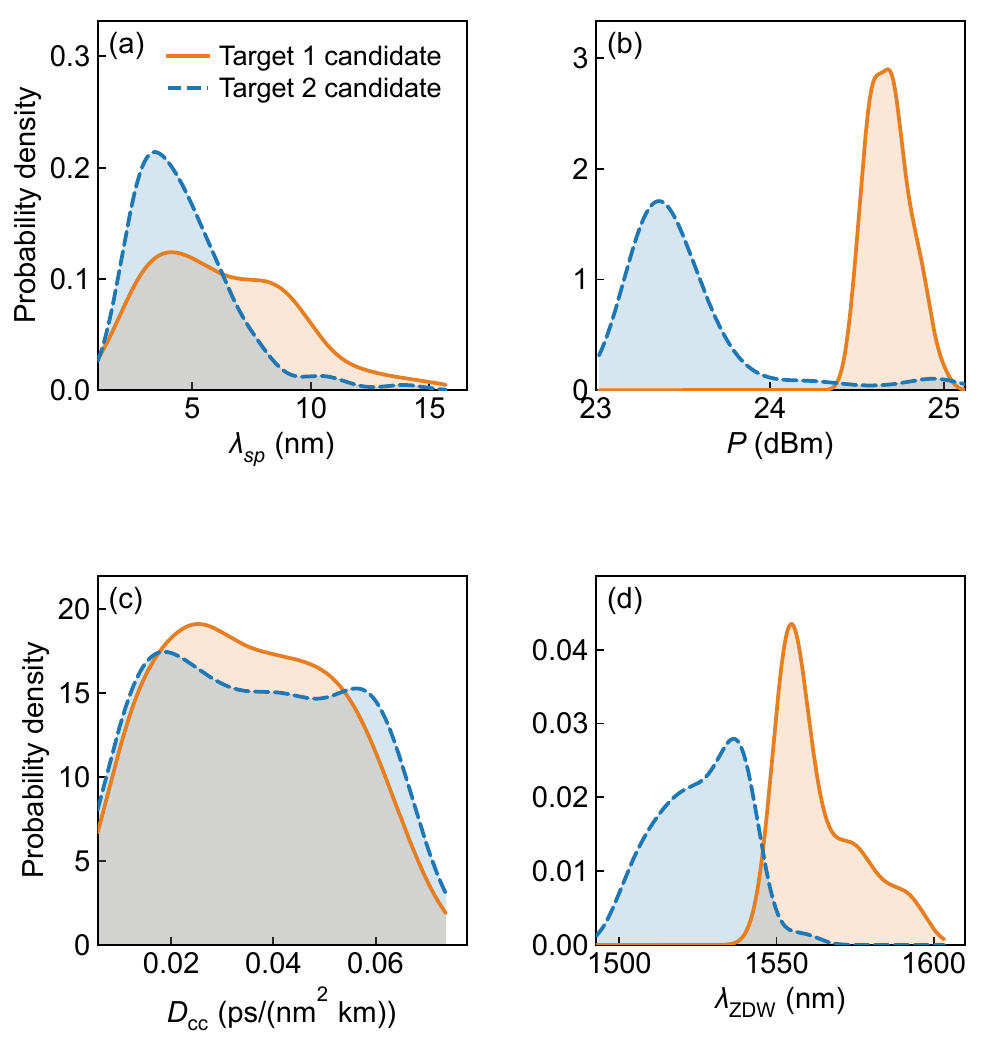}
    \caption{Input-parameter distributions of 100 local cluster samples for target 1 (in orange) and target 2 (in blue), as a function of (a) Pump-signal wavelength separation, (b) pump power, (c) dispersion slope $D_{cc}$ and (d) zero-dispersion wavelength $\lambda_{ZDW}$.}
    \label{fig:kmeans_input_distributions}
\end{figure}
To illustrate the selection of operating conditions,
we choose two targets: 1. $\mathrm{NF}=0.3~\mathrm{dB}$ and
$G_{\max}=20~\mathrm{dB}$ and 2. $\mathrm{NF}=1~\mathrm{dB}$ and
$G_{\max}=10~\mathrm{dB}$.
Figure~\ref{fig:kmeans_input_distributions} compares the
input-parameter distributions of 100 local candidates near the targets (target 1 in orange and target 2 in blue).
In Fig.~\ref{fig:kmeans_input_distributions}\,(a), the
pump-signal wavelength separation $\lambda_{sp}$ is
concentrated towards a smaller value, with the highest
density below 5~nm for both targets. This is not surprising, considering a higher gain is generally obtained when the pump-signal separation is low\,\cite{Xie2015MultiWavePSA}.
Figure~\ref{fig:kmeans_input_distributions}\,(b) again shows an expected trend of a dual-pump PSA with a positive correlation between the maximum gain and the pump power.
Fig.~\ref{fig:kmeans_input_distributions}\,(c) show a flat distribution profile for both targets with respect to the dispersion slope $D_{cc}$, indicating its low relative importance for the design of a high gain and low NF amplifier. 
Finally Fig.~\ref{fig:kmeans_input_distributions}\,(d) shows an important distinction between targets 1 and 2. According to this model, while achieving a moderate gain and NF (target 2) is possible through an operation in the anomalous dispersion regime ($\lambda_{ZDW}<1550$ nm), achieving a high gain and low NF (target 1) is almost always associated with an operation in the normal dispersion regime ($\lambda_{ZDW}>1550$ nm), and close to the signal wavelength (1550 nm).
Thus by clustering and rapidly identifying data points near the target region(s) we demonstrate an exploration of the parameter space for efficient design of the amplifier. 

\section{Conclusion \& Perspectives}\label{sec:con}
To conclude, in this work we developed an optimized deep-learning framework for rapid and accurate quantum NF calculation in a dual-pump fiber PSA. With the DNN, we demonstrated about 400 times speed up of the NF computation time. A prediction accuracy of more than 99\% was achieved. Using clustering we made exploratory searches over the amplifier parameter space, and retrieved parameters predicting about 20 dB maximum gain with a 0.3 dB NF. This calculation framework opens doors for rapid parameter space exploration of nonlinear fiber-optic systems in general and the design of high-gain and low-noise parametric amplifiers in particular. 

A future perspective may include a more accurate model development including larger number of waves in the NF calculation framework. Data-driven\,\cite{ermolaev2022data} and operator network\,\cite{sarkar2026learning} based approaches, are also interesting to understand the physics of noise dynamics in these systems. Another perspective of the present work is to explore inverse-design approaches based on tandem neural networks \cite{Xu2021Tandem} and conditional variational autoencoders (CVAEs) \cite{Tang2020CVAE} to infer input parameters that achieve specific gain and NF targets. 
Finally, these models are envisaged to be a stepping-stone towards a fully quantum NF evaluation model, enabling the design of next generation fiber-optic amplifiers.

\section{Appendix}\label{appendix}
This appendix summarizes the network architecture and hyperparameters used to train the surrogate model. Table~\ref{tab:raytune_search} lists the hyperparameter
search space explored using Ray Tune optimization. Ray Tune is used to coordinate the optimization trials, while the tree-structured Parzen estimator (TPE) implemented in Optuna proposes candidate hyperparameter configurations \cite{akiba2019optuna}. The asynchronous successive-halving algorithm (ASHA) is employed as the trial scheduler to terminate poorly performing configurations at an early stage \cite{li2020system}. Optuna exploits the results from previous trials to guide subsequent parameter suggestions, whereas ASHA reduces the computational resources allocated to unpromising configurations.

The hyperparameters search is configured for 100 trials, with a maximum of 500 epochs per trial. ASHA uses a 75-epoch grace period and a reduction factor of three. In addition, a trial is also stopped if its validation loss does not improve for 80 consecutive epochs. These trial-stopping rules operate separately from the cosine-annealing schedule, which controls the learning rate within each trial.

Table~\ref{tab:raytune_search} summarizes the search over learning rate, weight-decay coefficient, mini-batch size and network architecture.
Several predefined network architectures containing between four and six hidden layers are considered. The initial learning rate and weight-decay coefficient are sampled log-uniformly, whereas the remaining hyperparameters are selected from discrete sets. The activation function, optimizer, and loss function are kept fixed throughout the hyperparameter search.

Table~\ref{tab:ml_hyperparameters} reports the settings of the final optimized DNN.
The codes used for data generation, model
training and hyperparameter search, in this work are available
in the GitHub repository at \url{https://github.com/Yousra-bsr/Fiber_Amplifier_ML}.
To facilitate the practical use of the developed DNN, the repository also provides a ready-to-use prediction function, allowing users to directly predict NF and maximum gain from the four input parameters without requiring ML expertise or running numerical simulations.
\begin{center}
\begin{table}[htb!]
\centering
\caption{Hyperparameter search space.}
\label{tab:raytune_search}
\small
\setlength{\tabcolsep}{3pt}
\renewcommand{\arraystretch}{1.15}
\begin{tabular}{|p{0.41\textwidth}|p{0.50\textwidth}|}
\hline
\textbf{Hyperparameter} & \textbf{Search space} \\
\hline\hline
Hidden-layer configuration 
& 9 predefined architectures with 4-6 hidden layers \\
\hline
Initial learning rate
& $[1\times10^{-4},\,1\times10^{-3}]$ \\
\hline
Mini-batch size
& $\{64,128,256\}$ \\
\hline
Weight-decay coefficient
& $[10^{-6},\,10^{-4}]$ \\
\hline
\end{tabular}
\end{table}
\end{center}

\begin{center}
\begin{table}[htb!]
\centering
\caption{Network architectures and final DNN settings.
Hidden-layer widths are listed in order from input to output.}
\label{tab:ml_hyperparameters}
\small
\setlength{\tabcolsep}{2pt}
\renewcommand{\arraystretch}{1.15}
\begin{tabular}{|p{0.4\textwidth}|p{0.45\textwidth}|}
\hline
\textbf{Parameter} & \textbf{Value}  \\
\hline\hline
Main dataset size
& $300000$
\\
\hline
External dataset size
& $50000$
\\
\hline
Hidden-layer widths
&  $(512,512,256,256,256, 128)$
 \\
\hline
Initial learning rate
& $1.1\times 10^{-4}$
 \\
\hline
Minimum learning rate
& $10^{-6}$
\\
\hline
Weight-decay coefficient
& $1.78\times10^{-6}$
\\
\hline
Mini-batch size
& $128$
 \\
\hline
LeakyReLU negative slope
& $0.01$ \\
\hline
Dropout probability
& $0$\\
\hline
Maximum training epochs
& $500$ \\
\hline
Final early stopping patience
& $60$ epochs\\
\hline
\end{tabular}
\end{table}
\end{center}


\section*{Acknowledgment}




%

\end{document}